\documentclass{aa}  

\usepackage{xcolor}
\usepackage{graphicx}
\usepackage{txfonts}
\def\simless{\mathbin{\lower 3pt\hbox
     {$\rlap{\raise 5pt\hbox{$\char'074$}}\mathchar"7218$}}}   
\def\simmore{\mathbin{\lower 3pt\hbox
     {$\rlap{\raise 5pt\hbox{$\char'076$}}\mathchar"7218$}}}   

\def\msun{~{\rm M}_\odot}

\begin{document}

\title{The role of outflows in black-hole X-ray binaries}

\subtitle{}

\author{N. D. Kylafis\inst{1,2}
\and
P. Reig\inst{2,1}
}
\institute{University of Crete, Physics Department \& Institute of Theoretical and Computational Physics, 70013 Heraklion, Crete, Greece\\
\email{kylafis@physics.uoc.gr}  
\and   
{Institute of Astrophysics, Foundation for Research and Technology-Hellas, 71110 Heraklion, Crete, Greece\\
\email{pau@physics.uoc.gr}}}

\date{}

 
\abstract
{
The hot inner flow in black-hole X-ray binaries (BHBs) 
is not just a static corona rotating around the black hole, but  it must be partially outflowing. It is therefore a mildly relativistic ``outflowing corona''.
We have developed a model, in which Comptonization takes place in this outflowing corona. In all of our previous work, we assumed a rather high outflow speed of $0.8c$.
}
{
Here, we investigate whether an outflow with a significantly lower speed
can also reproduce the observations.  Thus, in this work we consider an outflow speed of $0.1c$ or less.
}
{
As in all of our previous work, 
we compute by Monte Carlo not only the emergent X-ray spectra, but also the time lags that are introduced to the
higher-energy photons with respect to the lower-energy ones 
by multiple scatterings. We also record the angle (with respect to the symmetry axis of the outflow) and the height at which photons escape.
} 
{
Our results are very similar to those of our previous work, with some small 
quantitative differences that can be easily explained. We are again able to reproduce quantitatively five observed correlations: a) the time lag as a function of Fourier frequency, b) the time lag as a function of photon energy, c) the time lag as a function of  
$\Gamma$, d) the time lag as a function of the cut-off energy in the spectrum, and e) the long-standing radio -- X-ray correlation.  All of them with only two parameters, which vary in the same ranges for all the correlations.
}
{
Our model does not require a compact, narrow relativistic jet, although its presence does not
affect the results.   The essential ingredient of our model is the
parabolic shape of the Comptonizing corona. The outflow speed plays a minor role.  Furthermore, the bottom of the outflow, in the hard state, looks like a ``slab'' to the incoming soft photons from the disk, and this can explain the observed X-ray polarization, which is along the outflow.  In the hard-intermediate state, we predict for GX 339-4 that the polarization will be perpendicular to the outflow.}

\keywords{accretion, accretion disks -- 
X-ray binaries: black holes -- 
jets -- 
X-ray spectra 
}

\authorrunning{Reig \& Kylafis 2023}

\titlerunning{The role of the outflow in BHBs }

   \maketitle
%

\section{Introduction}

It is typically accepted  that the broad picture of black-hole
X-ray binaries (BHBs) consists of an accretion flow (a geometrically thin, cool
outer disk and a geometrically thick, hot inner flow) and a narrow relativistic jet. 
The jet emits in the radio and the infrared, and the hot inner flow acts as a corona that
up-scatters soft disk photons to produce the hard X-rays.

The energy spectrum is nicely explained in the above picture as follows: a)
radio and at least part of the infrared come from the jet,  b) optical,
ultraviolet, and some soft X-rays come from the accretion disk,  and c) hard
X-rays come from the corona (for a review see Remillard \& McClintock 2006).  On the other hand, it is
well known that the spectra of BHBs are almost infinitely degenerate, in the
sense that if one is granted freedom on the geometry of the source, the
selection between thermal and non-thermal electrons, their energy distribution,
and the optical depth to electron scattering, almost any observed energy
spectrum can be fitted.

For energies above $\sim2$ keV, the harder photons are observed to lag with respect to softer
ones. These lags have been explained as the result of propagating fluctuations
in the hot inner flow \citep{nowak99c,kotov01,arevalo06,uttley11,uttley23}, light-travel
times in the outflow \citep{reig03}, the result of impulsive bremsstrahlung injection
occurring near the outer edge of the corona \citep{kroon16}, or as the evolution
time-scales of magnetic flares produced when magnetic loops inflate and detach
from the accretion disc \citep{poutanen99}. However, the observed correlation of
the lags with  the photon-number power-law index $\Gamma$ of the hard X-rays
\citep{kylafis18,reig18} and with the high-energy cutoff $E_c$ \citep{altamirano15},
suggest a common origin for the hard X-rays and the time lags. Likewise, the
fact that the time-lag -- $\Gamma$ correlation is inclination dependent
\citep{reig19} is also difficult to explain with the propagating-fluctuations model.

An important fact that has been neglected in the above picture is that the Bernoulli integral of the hot inner flow is positive \citep{blandford99}.  This means that the matter cannot fall into the black hole. Thus, part of the hot inner flow must escape as an outflow to leave the rest with a negative Bernoulli integral.
In other words, the hot inner flow is not just a static corona rotating around the black hole, but a wind-like, ``outflowing corona''.  

We have been promoting the idea that the above outflow from
the hot inner flow is the place where the X-ray
spectrum is shaped.  The reason is the following:  the outflow lies above and
below the hot inner flow. Thus, soft photons that are up-scattered in the hot
inner flow, must, before they escape, traverse the outflow, where they continue
the scattering process.  Since, after a few scatterings, the photons forget
their initial energy, it is the scattering in the outflow that determines
the emergent X-ray spectrum.

This picture has the advantage that it is very simple.  All it requires is a parabolic outflow in which Compton up-scattering of soft photons from the
accretion disk takes place.  No additional mechanism is required for the time
lags.  They come naturally with the Compton up-scattering.  The harder photons
are scattered more times than the softer ones, spend more time in the outflow,
and therefore come out later than the softer ones.  The size of the outflow and
its optical depth determine the magnitude of the time lags.  Also, the fact that
the time lags and the hard X-ray spectrum are produced by the same process
(Comptonization), means that it is not surprising that the two are correlated
\citep{reig15,kylafis18}. 
Furthermore, the Comptonized X-ray spectra that come out of the outflow are
anisotropic, because of the shape of the outflow (parabolic) and the outflow speed ($v_0=0.8c$).   The harder photons come
out mainly along the outflow (large optical depth) and the softer ones mainly
perpendicular to it (smaller optical depth).  This, then, naturally explains the
inclination dependence of the time-lag -- $\Gamma$ correlation \citep{reig19}.

The parabolic outflow also emits radio waves.  In fact, the whole spectrum from
radio to hard X-rays can be explained in the outflow model \citep{markoff01,giannios05}.  
In
addition, since the same electrons do the Compton up-scattering and the radio
emission by synchrotron, it is not surprising that the radio flux correlates
with the X-ray flux \citep{corbel13, kylafis23}.

In our previous works, we used the word jet to refer to the outflow.  In
the past, anything outflowing was called "jet".  This name now appears
inappropriate.  Our work reveals that a narrow relativistic jet is not required
in this picture.  There is no problem if it exists, but it is not necessary. 

Here, we want to distinguish between the 
jet in BHXRBs, which is narrow (a few  $R_g$ at its base, $R_g$ is the gravitational radius) and relativistic ($v_0
\simmore 0.7 c$) and the outflow in BHXRBs, which is wind-like, broad ($10
- 10^3 R_g$ at its base), and non relativistic.  The outflow speed cannot be lower than the local escape speed, which is $v_{esc}(R)/c = \sqrt{2R_g/R}$, with $R$ the radial distance.  This means that the outflow speed $v_0(R)$ is $> 0.45 c$ at $R= 10 R_g$ and $> 0.045 c$ at $R=10^3 R_g$.
Since in our previous work we used $v_0 = 0.8 c$, which
for a wide outflow (say $10^2 - 10^3 R_g$)  implies a huge mass-outflow rate, unless the
matter consists of electron - positron pairs (not likely in BHBs), we need to demonstrate that an outflow with significantly lower speed
reproduces the above correlations.  This is what we demonstrate in the present paper, but with $v(R) =$ constant $=0.1 c=v_0$ and not a function of $R$.  The reason for this is because we do not know the mass-outflow rate at every radius $R$, in order to compute the density in the outflow at a given height as a function of radius.  Thus, we feel that a demonstrative calculation, with a low constant outflow speed, will suffice.  We comment about this in Sect.~\ref{incl} and ~\ref{summary}.

 A detailed description of our model is given in Appendix A, where we  also describe how our Monte Carlo code works. In Sect.~\ref{results},  we discuss our results, and in Sect.~\ref{summary} we give a summary and present our conclusions.


\section{Results}
\label{results}

In the next sections we present the observational results that our model is
capable of reproducing. These results are to be compared with those published
over the years in \citet{reig03}, \citet{kylafis08}, \citet{reig15},
\citet{kylafis18}, \citet{reig18}, \citet{reig19}, \citet{kylafis20}, and
\citet{reig21}. Our goal is to demonstrate that a non-relativistic, wind-like outflow (i.e., $v_0=0.1c$) can reproduce our previous results (where a mildly 
relativistic outflow with $v_0=0.8c$ was used), which in turn, explain many observations and correlations. In this work, we have assumed that the observer sees the system at an intermediate inclination angle. Hence we combined all the escaping photons with directional cosines in the range $0.2 \leq \cos \theta \leq 0.6$. A full description of the meaning of the model parameters is given in Appendix A.

\begin{figure}
\centering
\includegraphics[width=8cm]{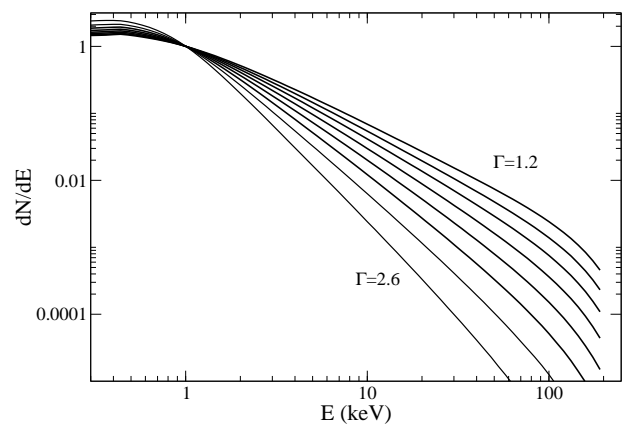}
\caption{Energy spectra for different optical depths: from top to bottom
$\tau_\parallel=4.5, 4, 3.5, 3, 2.5, 2, 1.5, 1$. The width at the base of
the outflow was $R_0=400\,R_g$ and the Lorentz factor of the coasting electrons was $\gamma_0= 1.14$.  The spectra have been normalized by the flux at 1 keV.} 
\label{jetspec}
\end{figure}

\subsection{Energy spectra}
\label{energy}

The observed X-ray spectral continuum (0.1--200 keV) of BHBs
consists of a soft thermal component and a hard non-thermal component.  The
thermal component is modeled as a  multi-temperature black-body  and  dominates the
spectrum below $\sim 2$ keV \citep{mitsuda84,merloni00}. Its origin is
attributed to a  geometrically thin, optically thick accretion disk \citep{shakura73}. The
non-thermal component is well described by a power law with a high-energy
exponential cutoff. This component is believed to be the result of
Comptonization of low-energy photons from the accretion disk \citep{sunyaev79}, by energetic electrons in a configuration that is still under debate.

During X-ray outbursts, BHBs go through different spectral states
\citep{mclintock06,belloni10}, of which the two main ones are the soft and hard
states \citep{done07}. Intermediate states (hard-intermediate, HIMS, and soft-intermediate, SIMS) occur when the source transits
from these basic states. Each state is characterized by a different contribution
of the thermal and power-law components.  In the soft state (SS), the thermal
black-body component dominates the energy spectrum, with no or very weak power-law
emission \citep{remillard06,dexter12}. In this state the radio emission is
quenched. In the hard state (HS), the soft component is weak or absent, whereas the power-law extends to a  hundred keV or more with
photon-number index $\Gamma$ in the range $1.2-2$. In the HIMS, the power-law component is still present, albeit with a steeper slope (i.e., larger $\Gamma$) than in the HS, while the black-body component starts to appear \citep{mclintock06,castro14}, if it was not already there in the HS. In the HS and the HIMS, the systems shine bright in the radio band.

In our model, the Comptonizing medium is the outflow, which extends laterally to relatively large distance (few hundred of $R_g$ at its base) from the black
hole. Therefore, our model is relevant for the HS and HIMS and it should be able to reproduce a power-law with photon index in the range $1.2-2.6$, as measured in the observations.  

The main parameter that affects the slope of the continuum is the optical depth $\tau_\parallel$ ($\tau_{\perp}$ is uinquely determined given $\tau_\parallel$ and $R_0$), as it is a parameter directly related to the number of scatterings in the outflow. Indeed, we can reproduce the observed range of photon index $\Gamma$ in the HS
and the HIMS by changing the optical depth
$\tau_\parallel$. Figure~\ref{jetspec} shows the resulting energy spectra of our
model. Each spectrum corresponds to a model with $R_0 = 400\, R_g$, Lorentz $\gamma_0 = 1.14$, and variable
$\tau_\parallel$ between 1 and 4.5, while the rest of the parameters are kept
fixed at the reference values given in Table~\ref{modpar}. The spectra have been normalized by the flux at 1 keV. The non-relativistic
outflow ($v_0=0.1 c$) requires lower values of the optical depth to reproduce the photon index
of the HS and HIMS, compared to the mildly relativistic outflow that we used before ($v_0=0.8 c$), for which $\tau_\parallel$
ranged between 2 and 11 \citep{reig03,kylafis18}.

\begin{figure}
\centering
\includegraphics[width=8cm]{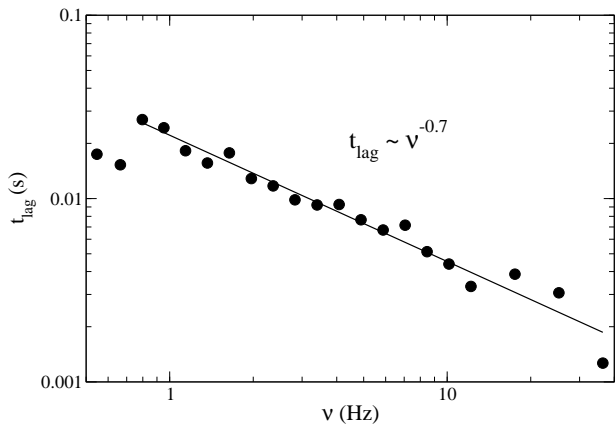}
\caption{Time lag as a function of Fourier frequency. The model shown corresponds to
$\tau_\parallel=3$, $R_0=400\,R_g$, and $\gamma_0 = 1.14$.}   
\label{jetlag}
\end{figure}

\begin{figure}
\centering
\includegraphics[width=8cm]{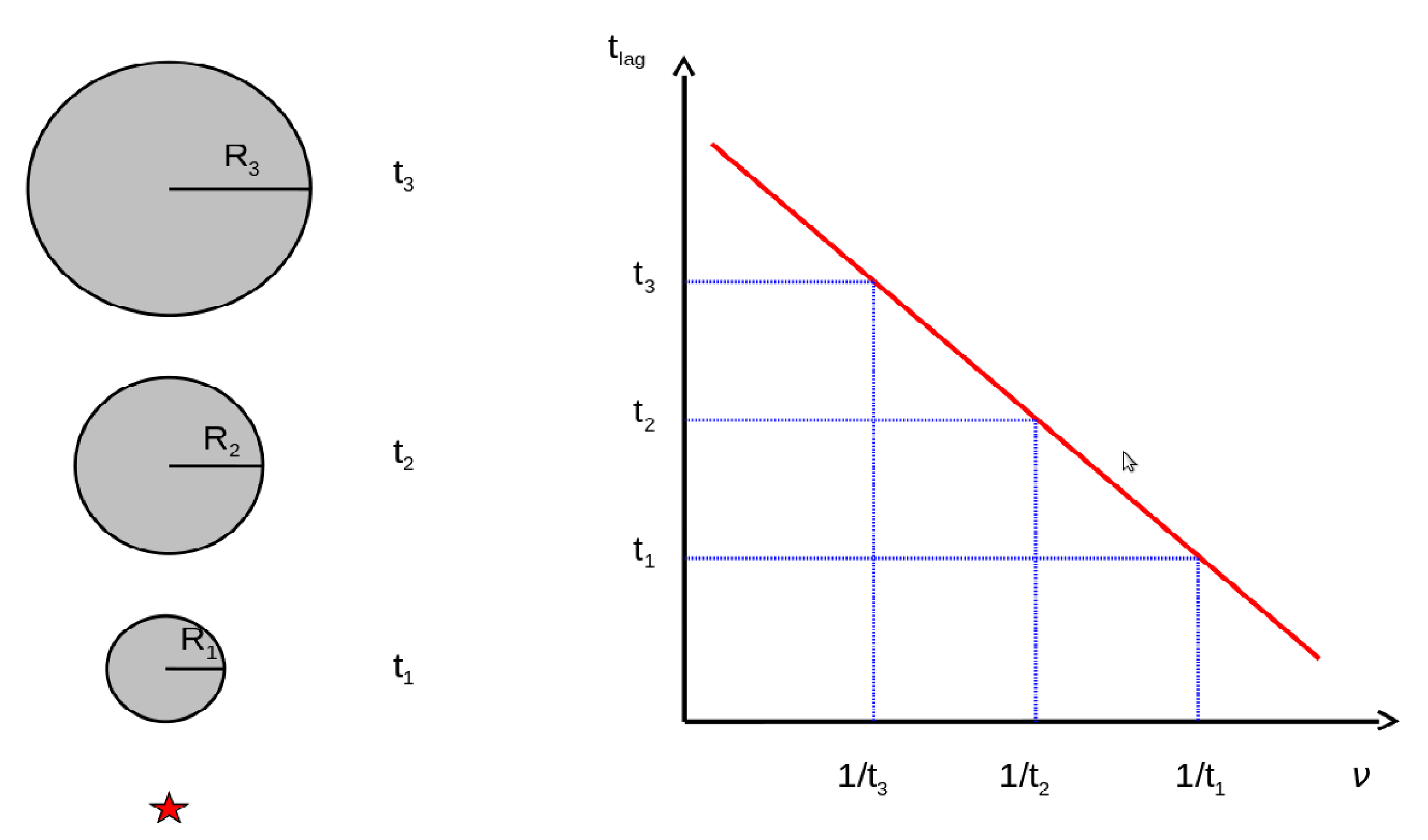}
\caption{Explanation of the frequency dependence of time lags.}  
\label{jetlag_model}
\end{figure}

\subsection{Lag-frequency correlation}
\label{lags}

When two light curves obtained at two separate energy bands (say, a hard and a soft) are
cross-correlated, lags are observed between the hard and soft bands. Positive or hard
lags mean that the hard photons lag the soft photons. This is always the case
at energies above $\sim 2$ keV, that is, at energies where the power-law
component, typically associated with Comptonization, dominates. The magnitude of
this lag strongly depends on Fourier frequency and on the energy bands
considered \citep{miyamoto88,vaughan97,cui97,poutanen01,pottschmidt03}, but typically it
is in the range of 1--100 ms. A different type of lags are the so-called
reverberation or soft lags, where the soft photons (say, around 0.5 keV) lag the hard ones.  These can 
result from the delay between the hard photons that hit the accretion disk, are absorbed there, and are 
emitted as soft photons, and the hard photons that reach the observer directly, before the soft photons
\citep{uttley14}. In this case the soft band is
defined at energies below $\sim 1$ keV, as it comes from the disk \citep{uttley11}. Reverberation lags are typically negative, that
is, soft photons lag hard photons \citep{kara19}. This is explained by the extra path that the absorbed hard X-rays have to travel. In this work we deal with positive lags only. 

The positive lags, calculated at energies above $\sim 2$ keV, can naturally be attributed to inverse Comptonization. In order to acquire their energy, harder
photons scatter more times than less hard photons, hence they stay longer in the Comptonizing medium before they escape. In this context, time lags simply signify the difference in light-travel time of photons within the Comptonizing region. A different explanation of the time lags was offered by \citet[][see also \citealt{kotov01} and \citealt{arevalo06}]{lyubarskii97}. In this model, the lags result from viscous
propagation of mass-accretion fluctuations within the inner regions of the accretion flow.

Observations of time-lags in the hard state show that they roughly
follow a power-law dependence on Fourier frequency of the form $t_{\rm lag} \propto \nu^{-0.8\pm0.1}$
\citep{crary98,nowak99a,cassatella12}. 
Our model reproduces quantitatively this nearly $1/\nu$ dependence of the time-lags on Fourier frequency, and this is shown in Fig.~\ref{jetlag}. 
The time of flight of all escaping photons was recorded in $N_{\rm bins} = 8192$ time bins of duration 1/64 s each. This time was computed by adding up the path lengths traveled by each photon and dividing by the speed of light. Then, we considered the light curves of two energy bands: soft or reference band (2 - 6 keV) and hard (6 - 15 keV). We  identified the phase lag $\phi$ 
between the signals of the two bands as the phase of the complex cross-vector, and from it the corresponding time lag $\tau=\phi/2\pi\nu$
as a function of Fourier frequency $\nu$.

The black filled circles in Fig.~\ref{jetlag} correspond to a model with $\tau_{\parallel}=3$, $R_0=400\,R_g$, and $\gamma_0= 1.14$. The values of these parameters are not crucial, as long as $\tau_{\parallel}$ is not too large.  The case of large  $\tau_{\parallel}$ will be explored in a subsequent paper (Reig, Kylafis, \& Pe'er, in preparation).  It is straight forward to explain qualitatively the above approximately $1/\nu$ dependence of the time-lags, because Compton scattering acts like a filter that cuts off the high frequencies.  We explain this below.

For the sake of this argument, let's think of the outflow as a series of spheres with radii $R_1 < R_2 < R_3 <...$, one on top of the other, with the smaller one at the bottom  (see Fig.~\ref{jetlag_model}).  This is a discrete visualization of the parabolic outflow with radius $R(z) = R_0 (z/z_0)^{1/2}$, where $z_0$ is the height above the black hole at which the outflow starts.  If $\tau_{\parallel}$ is of order unity, then soft photons from the accretion disk, that enter the outflow from below, can scatter in any of the above spheres. The source of soft input photons is indicated in Fig. 3 by a star. Consider soft photons that have an intrinsic variability with period $P$, i.e., frequency $\nu= 1/P$, and scatter in one of the spheres.  If they
scatter in the sphere with radius $R_1$, this variability will be significantly reduced if the time delay due to Compton scattering, $t_1 = R_1/c$, is comparable to or larger than $P$.  This means that all frequencies larger than $1/t_1 = c/R_1$ will be essentially washed out.  If the scattering occurs in the sphere with radius $R_2$, with typical time delay $t_2 = R_2/c$, then all frequencies larger than $1/t_2 = c/R_2$ will be washed out, and so on.  In other words, if the time delay due to Compton scattering is $t$, then all frequencies of variability in the input photons larger than $1/t$ will be washed out.  The Monte Carlo Comptonization in a parabolic outflow reproduces the $\nu^{-\alpha}$ dependence, with $\alpha=0.7-0.9$.  If the outflow is not parabolic, then Comptonization does not reproduce this frequency dependence.  Such outflows will be examined in a subsequent paper and they seem to be related with the outlier sources, i.e., the sources that do not obey the regular radio -- X-ray correlation \citep{gallo03,corbel03,kylafis23}.

\begin{figure}
\centering
\includegraphics[width=8cm]{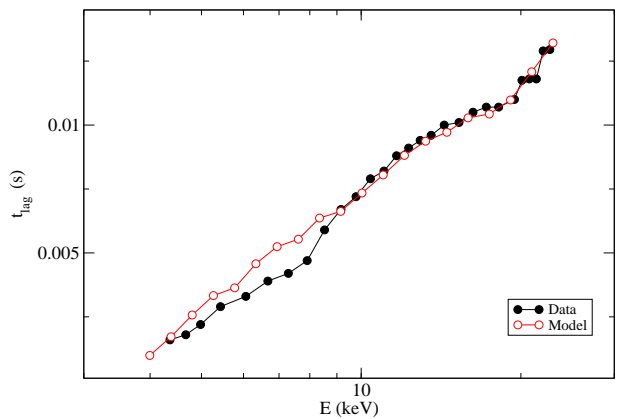}
\caption{Lag as a function of energy. Black filled circles represent the observations for Cyg X--1. The model shown as red empty circles corresponds to $\tau_{\parallel}=2$, $R_0=180\,R_g$, and $\gamma_0 = 1.14$. 
}. 
\label{lagen}
\end{figure}

\subsection{Lag-energy correlation}

In the previous section, we showed that the hard lags observed in the HS of BHBs
can be due to the fact that the harder photons have undergone more scatterings inside the outflow (the
region where the energetic electrons reside) than lower-energy photons. Also, the energy of the photons changes with every scattering, with an increase in energy, on average. Therefore, both the final amplitude
of the lags and the energy with which the photons escape depend on the number of scatterings. Comptonization models predict a log-linear energy dependence in the
lags, approximately what is observed \citep{nowak99a,kotov01,stevens16}. The
slope of the relation depends on the range of Fourier frequencies considered to compute the average lag, becoming
flatter as the frequency increases \citep{kotov01,uttley11} 
Observations with good signal-to-noise
show that the relation may be more complex than a simple log-linear law, with
some bumps or breaks, around the energy of the iron line at 6.4 keV
\citep{kotov01,stevens16}.

Figure~\ref{lagen} shows the energy dependence of the time lag, as computed from our model with $R_0=180R_g$, $\tau_{\parallel}=2$, and $\gamma_0=1.14$. The data
are from \citet{kotov01} for Cyg X--1. The reference band is 2.7--4 keV and the
frequency range of the lag calculation is 0.05--5 Hz, to approximately match the values used by \citet{kotov01}. 
In general terms, our model reproduces quite well the approximately linear relation between $t_{\rm lag}$ and $\log E$.

\begin{figure}
\centering
\resizebox{\hsize}{!}{\includegraphics{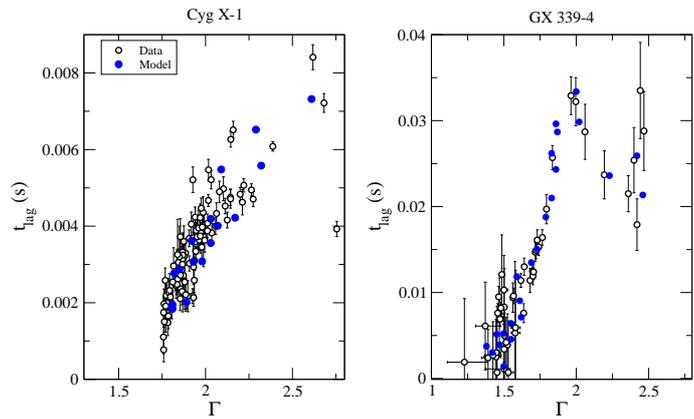}}
\caption{Correlation between time lag and photon index $\Gamma$ for Cyg X--1 (left) and
GX\,339--4 (right). The data for Cyg X--1 come from \citet[][see also \citealt{kylafis08}]{pottschmidt03} and
for GX\,339--4 from \citet{kylafis18}. Black empty circles represent the
observations and the filled blue circles correspond to the
models, which are produced with $\gamma_0 = 1.14$ and the values of $\tau_{\parallel}$ and $R_0$ that are shown in Fig.~\ref{tau-R0}}.
\label{lagamma}
\end{figure}

\subsection{Lag - $\Gamma$ correlation}
\label{lagindex}

Numerous studies have shown that the spectral (e.g., photon index $\Gamma$, cutoff energy $E_c$)
and timing (characteristic frequency of the broad-band noise and QPOs, time or
phase lags) quantities in BHBs display tight correlations
\citep{dimatteo99,pottschmidt03,kalemci03,kalemci05,shaposhnikov09,stiele13,shidatsu14,
grinberg14,kalamkar15,altamirano15,kylafis18,reig18,reig19,karpouzas21,mendez22}.
These correlations represent convincing evidence that the timing and
spectral properties of the sources are closely linked.


We have demonstrated that our outflow model is able to reproduce the observed
correlation between time lag and $\Gamma$, not just for a specific source CygX--1 \citep{kylafis08} or
GX\,339--4 \citep{kylafis18}, but in general for
BHBs as a group \citep{reig18}.  All these correlations were
reproduced varying only the optical depth $\tau_\parallel$ and the width at the
base of the outflow $R_0$. In those models, the Lorentz factor was $\gamma_0=2.24$,
which resulted from a high outflow speed $v_0 =0.8c$ and a moderate
perpendicular component $v_{\perp}=0.4c$. In this work, we show that the outflow
speed is not a critical parameter of the model and that a non-relativistic
outflow with $v_0 = 0.1 c$ can also reproduce the observations.  We take $v_{\perp} = 0.47 c$, so that
$\gamma_0 =1.14$.  

In Fig.\ref{lagamma} we show the  $t_{\rm lag}-\Gamma$ correlation for Cyg X--1
(left panel) and for GX\,339--4 (right panel). The black empty circles represent the
observations and the blue filled circles our models. These figures are to be
compared with Fig. 2 in \citet{kylafis08} and Fig. 2 in \citet{kylafis18}. The
energy and frequency ranges used to compute the lags are the same as in the above
references, namely between the bands $2 - 4$ keV and $8 - 14$ keV in the frequency
range $3.2 - 10$ Hz for Cyg X--1, and between $2 - 6$ keV and $9 - 15$ keV in the $0.05 - 5$ Hz range for GX 339--4.
This comparison reveals that even if the outflow speed is reduced
from $0.8c$ to $0.1c$, the range of variability of the optical depth $\tau_{\parallel}$ and radius $R_0$
does not change significantly. At the lower outflow speed ($v_0=0.1 c$), the fits require slightly
smaller optical depth and larger radius, $1 \simless \tau_{\parallel} \simless 6$ and $50\, R_g
\simless R_0 < 1500\, R_g$, compared to $2 \simless \tau_{\parallel} \simless 11$ and $30\, R_g \simless R_0 \simless 600\, R_g$ for the $v_0=0.8 c$ case.

\begin{figure}
\centering
\resizebox{\hsize}{!}{\includegraphics{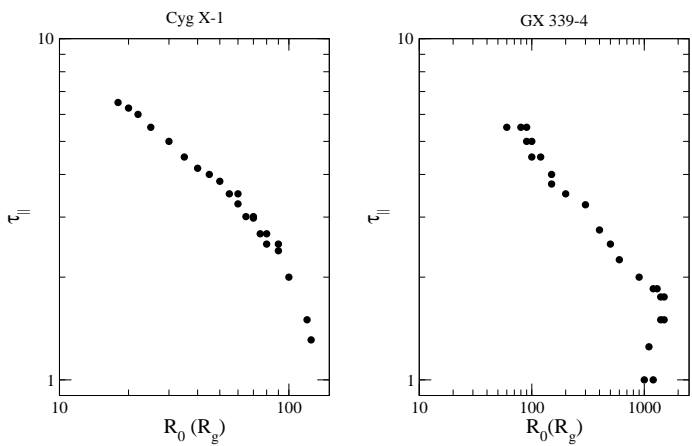}}
\caption{Correlation between the optical depth $\tau_{\parallel}$ and the radius $R_0$ of the outflow at
its base for Cyg X--1 (left) and
GX\,339--4 (right). The values of the parameters $R_0$ and $\tau_{\parallel}$ 
are those that were used in the models that produced Fig.~\ref{lagamma}.   }
\label{tau-R0}
\end{figure}

\subsection{$\tau_{\parallel}$ - $R_0$ correlation}

One remarkable outcome of our model is the correlation that we have found between the optical depth $\tau_{\parallel}$
and the radius $R_0$ at the base of the outflow
\citep{kylafis08,kylafis18,reig18,reig19}. As we mentioned before, we can
reproduce a number of observations and correlations by changing only the two basic parameters $\tau_{\parallel}$
and $R_0$. Surprisingly,  these two parameters do not vary independently of one another,
but in a correlated manner, following a power law $\tau_{\parallel} \propto
R_0^{-\beta}$, in the HS of BHBs. Thus, for the modeling of the HS of BHBs we basically need one parameter, not two.
Figure~\ref{tau-R0} shows this correlation. The points
correspond to the models that were used for the explanation of the $t_{\rm lag}-\Gamma$ correlations
shown in Fig.~\ref{lagamma}. The correlations break down when the source enters the
HIMS. The index $\beta$ is not unique for all BHBs, but different sources
display different values of $\beta$. For example, for Cyg X--1, $\beta=0.69$
and for GX\, 339--4, $\beta=0.38$. This is to be expected since the amplitudes of
the lags of the two sources are significantly different.

\begin{figure}
\centering
\resizebox{\hsize}{!}{\includegraphics{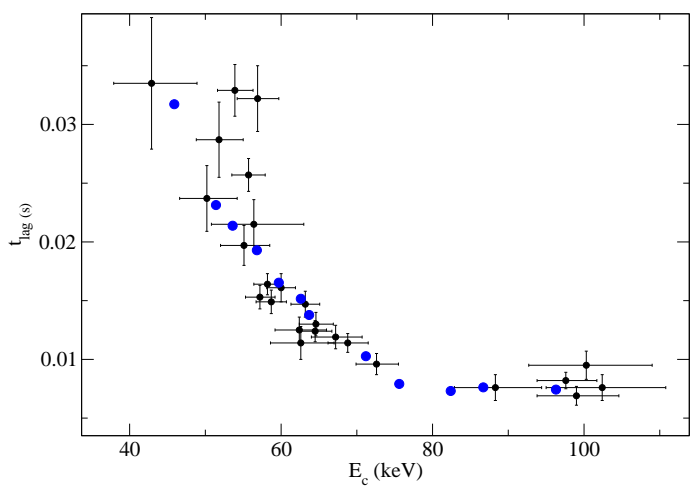}}
\caption{Relationship between the time lag and cut-off energy. Black circles
represent the observations for the BHB GX\,339--4. The blue circles
correspond to our models. The values of the parameters $R_0$ and $\tau_{\parallel}$ are the same as in the models that were used in Fig.~\ref{lagamma}.
}
\label{cutoff}
\end{figure}

\subsection{Lag -- cut-off energy correlation}

Another interesting result observed in the BHB GX\,339--4 is the correlated
evolution of the cut-off energy $E_c$ (in its energy spectrum) and the phase-lag $\phi_{\rm lag}$ of harder photons compared to less hard ones, as the source progresses
along the HS \citep{motta09,altamirano15}. As the X-ray flux increases, the cut-off
energy decreases and the amplitude of the time lag increases \citep[see Fig. 7
in][]{altamirano15}. We showed in Sect.~\ref{energy} that higher optical depths $\tau_{\parallel}$
result in harder spectra (the photons are scattered more and gain more energy
from the energetic electrons of the outflow). Likewise, by increasing $R_0$, the
spectrum also hardens, because larger $R_0$ translates to  larger Thomson
optical depth $\tau_{\perp}$ perpendicular to the axis of the outflow and therefore more
scatterings of the photons. For the same reason, the amplitude of the lags increases when the
optical depth and/or the radius of the outflow increase. The cut-off energy
$E_c$ has a weak dependence on  $\tau _{\parallel}$ and $R_0$, but a
rather strong dependence on $v_{\perp}$. This is because the cut-off is mainly
determined by the energetics of the electrons and $v_{\perp}$ determines the
maximum energy gain of the soft photons \citep{giannios04,giannios05}. 

In \citet{reig15}, we reproduced the correlation between the cut-off energy $E_c$ and  the phase lag $\phi_{\rm lag}$
 observed in the BHB GX\,339--4 by changing $\tau_{\parallel}$, or
$v_{\perp}$, or both in a correlated way. In that work, we
kept $R_0$ fixed at 100 $R_g$. However, as the outburst progresses (i.e., as the
X-ray flux increases), we expect that both $\tau_{\parallel}$ and $R_0$ will vary, as we have already shown in Sect.~\ref{lagindex}. Our model requires
that as the source moves from the HS to the HIMS, $\tau_{\parallel}$ decreases
and the $R_0$ increases.  

Here we reproduce again the correlation between lag and cut-off energy in a slightly different form (Fig.~\ref{cutoff}). We use time lag instead of phase lag and the results of our own analysis instead of data from \citet{motta09} and \citet{altamirano15}. The details of the analysis can be found in \citet{reig18}. We use data from the 2006 outburst of GX 339--4. In Fig.~\ref{cutoff}, black filled circles represent the
observations and blue filled circles the results of our models. We emphasize that the models displayed in Fig.~\ref{cutoff} are the same models (i.e., the same combinations of $\tau_{\parallel}$ and $R_0$) that reproduced the $t_{\rm lag}$ -- $\Gamma$ correlation of
Fig.~\ref{lagamma}, simply adjusting slightly the $v_{\perp}$. The values of $v_{\perp}$ that reproduce the observations vary in the range $0.41c - 0.46c$.

\begin{figure}
\centering
\resizebox{\hsize}{!}{\includegraphics{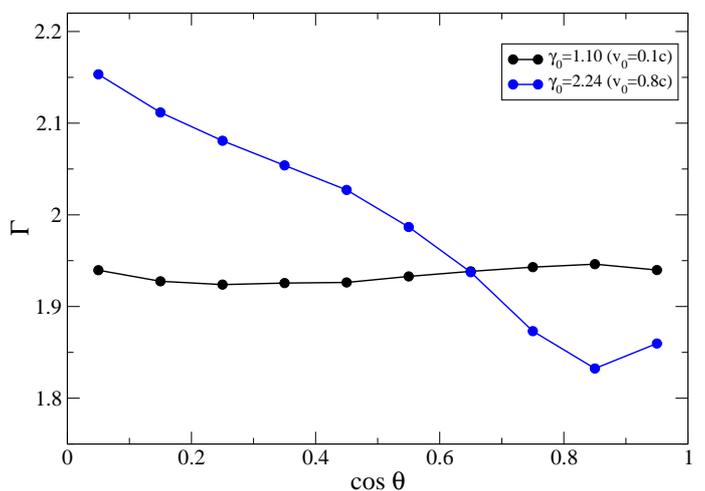}}
\caption{Photon index as a function of escaping direction. The blue dots are for $v_0=0.8 c$ and the black dots for $v_0= 0.1 c$.  In the inset, we show the corresponding Lorentz $\gamma_0$.  }
\label{angle-index}
\end{figure}

\subsection{Inclination effects}
\label{incl}
In \citet{reig18}, we investigated the correlation between $t_{\rm lag}$ and $\Gamma$ for a large number of BHBs. The data in the correlation showed a
large scatter that was explained as an inclination effect \citep{reig19}.
Systems seen at low inclination exhibit a stronger correlation. In
high-inclination systems, the correlation is rather weak. We also showed that low-
and intermediate-inclination systems tend to have harder spectra \citep[see
also][]{reig03} and a larger amplitude of time lags. The explanation is that,
in a mildly relativistic outflow, the high bulk speed of the electrons provides a boost on
the photons that makes them scatter preferentially in the forward direction.
These photons travel longer distances and suffer more scatterings than photons
that escape perpendicularly to the outflow axis. Therefore, photons that escape at
small to moderate angles $\theta$ with respect to the outflow axis lead to harder
spectra (smaller photon index), because they undergo more scatterings and longer
lags because they travel longer distances.

Figure~\ref{angle-index} shows the dependence of the photon index on the
escaping angle for the case of a mildly relativistic outflow with $v_0=0.8c$ (blue dots) 
and a
non-relativistic outflow with $v_0=0.1c$ (black dots). 
Here $\theta$ is the angle between the
observer and the outflow axis (i.e., photons with $\cos \theta \sim 1$ escape
along the axis, while $\cos \theta \sim 0$ escape perpendicular to the outflow
axis). The parameters of the models shown are $\tau_{\parallel}=5.5$ and $R_0=140 \,
R_g$ for the mildly relativistic outflow and $\tau_{\parallel}=3.5$ and $R_0=150 \, R_g$ for
the non-relativistic outflow. 

As we indicated in the Introduction, our model with constant $v_0 = 0.1 c$ is only demonstrative because in reality the escape speed is not expected to be
constant, but a function of 
$R$, and it ranges from $0.58 c$ at $R=6R_g$ to $0.045 c$ at $R=10^3 R_g$.  Thus, the real outflow has a central fast part and a progressively slower outer part.  Without this variable $v_0(R)$ it is not possible to compute accurately the spectra as a function of inclination.

\begin{figure}
\centering
\resizebox{\hsize}{!}{\includegraphics{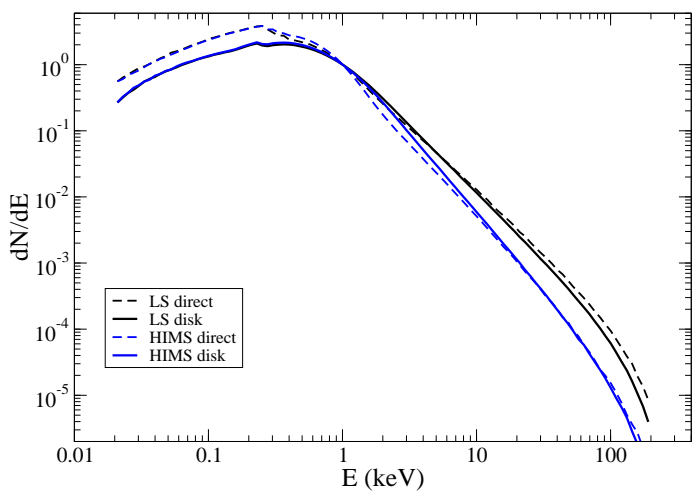}}
\caption{Comparison of the “direct” and “disk” spectra. The direct spectrum is the spectrum seen by observers at infinity at an inclination $\theta\sim40^{\circ}$ and the “disk” spectrum is the spectrum of the photons that are back-scattered and illuminate the disk. The parameters of the models used in this figure for the LS are $\tau_{\parallel}=2.5$, $R_0 = 500$ $R_g$, and $\gamma_0=1.14$ and for the HIMS $\tau_{\parallel}=1.5$, $R_0 = 1400$ $ R_g$, and $\gamma_0=1.14$.}
\label{disk-jet_spectrum}
\end{figure}

\subsection{Disk illumination}
\label{hdis}

In our model, soft photons are injected isotropically upward at the base of the outflow.  Depending on $\tau_{\parallel}$ and the initial direction, photons may 
escape unscattered or after a number of scatterings.  After escape, we record
their energy, angle of escape (with respect to the
outflow axis), height from the black hole, and travel time in the outflow. Hence,
the code also computes the back-scattered photons, i.e., photons with escaping angle $\theta > \pi/2$. Some of these photons will
be absorbed by the accretion disk and others will be reflected by it. The code,
in its current version,
does not account for reflection, but we can measure the number of photons that
hit the accretion disk. In other words, we can compute the irradiation spectrum.

In \citet{reig21}, we showed that the fraction of back-scattered photons
increases as the Lorentz factor $\gamma_0$ decreases. The reason is
that in non relativistic outflows the scattering is nearly isotropic, whereas when the outflow velocity is high, there is a strong forward boost, as explained in the previous section.
Although in our model photons can interact with
electrons anywhere in the outflow, most of the scatterings occur close to the
black hole, not far away from the base of the outflow if the outflow velocity is low. 

Fig.~\ref{disk-jet_spectrum} compares the energy distribution of the photons that illuminate the disk, that is the “disk” spectrum (solid line), with the “direct” spectrum of the photons from the outflow, 
seen by observers for typical LS and HIMS spectra. 
Unlike the relativistic case with $v_0=0.8c$, where the ``disk'' spectrum was much
softer than the ``direct'' one \citep[see Fig. 3 of][]{reig21},
here the ``disk'' and the ``direct'' spectra are very similar. 
The spectra, both ``direct'' and ``disk'', are slightly softer in the HIMS than in the HS, in agreement with observations.


Another way to investigate the irradiation of the accretion disk by the primary source is by computing the
emissivity profile, which is the radially dependent flux irradiating the disk by the source. For a standard Shakura-Sunyaev accretion disk, it is parameterized as a power law $F(r) \propto r^{-q}$, with $q$ the emissivity index. The standard behavior is $q=3$ \citep{dauser13}.
To compute the emissivity index, we divided the accretion disk into radial zones and computed the number of photons per unit area that irradiate a given zone. For the sake of the computation and in order to compare with the lamppost geometry, we  collapsed the outflow to $R_0=1\,R_g$. We find that for $R_{\rm disk} \simmore 10\, R_g$, the radial dependence of the irradiated flux on the disk does follow a power-law with $q=3$, that is, the expected value for a standard Shakura-Sunyaev disk\footnote{When relativistic effects are taken into account, steeper profiles are found in the inner parts of the accretion disk \citep{dauser22} and a broken power law is used instead \citep{bambi21}. We note that because our model deals with broad outflows, relativistic effects are not expected to play any significant role in our results.}.

\begin{figure}
\centering
\resizebox{\hsize}{!}{\includegraphics{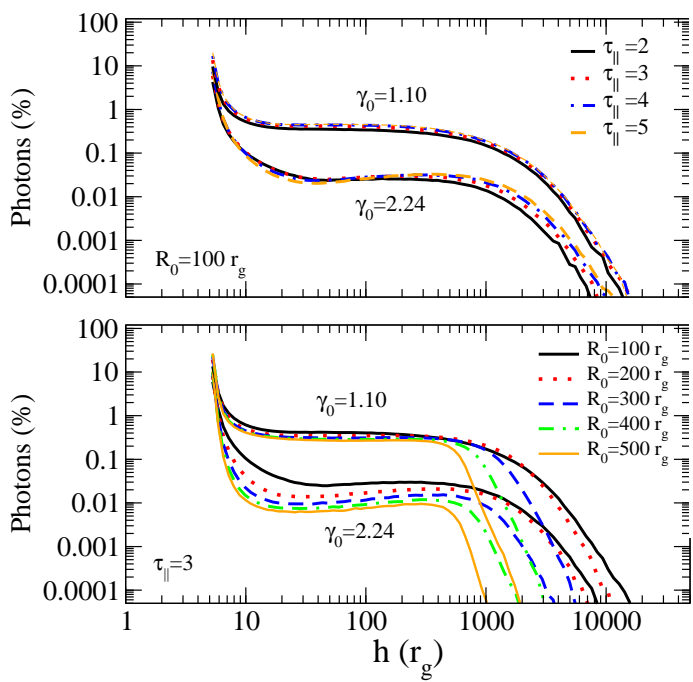}}
\caption{Fraction of photons that hit the accretion disk over the total number of photons as a function of distance $h$
from the black hole at which they escape, for different
values of the optical depth $\tau_{\parallel}$ and radius $R_0$ of the outflow. The distribution is shown
for a mildly relativistic outflow ($\gamma_0=2.24$, i.e., $v_0=0.8c$) and non relativistic outflow
($\gamma_0=1.10$, i.e., $v_0=0.1c$).}
\label{height}
\end{figure}

Fig.~\ref{height} shows the distribution of distances $h$ from the black hole from
which the photons that hit the accretion disk escape. We show this distribution
for different values of $\tau_{\parallel}$ (top panel) and $R_0$ (bottom panel)
and two Lorentz factors $\gamma_0=1.10$ and $\gamma_0=2.24$ (i.e., $v_0=0.1 c$ and $v_0=0.8c$, respectively). As expected, most
of the photons escape within a few $R_g$, but a significant fraction of the
photons that hit the accretion disk also escape at tens to few hundreds of
$R_g$.  This figure also confirms the fact that disk illumination increases as the outflow speed decreases.
The vertical axis in Fig.~\ref{height} is reminiscent of the reflection fraction, defined as  the number of emitted photons of the primary source which 
hit the accretion disk $N_{\rm AD}$ over the number of photons escaping to infinity $N_{\infty}$ \citep{dauser14,dauser16}. \footnote{Here we have employed $N_{\rm AD}/N_{\rm phot}$, instead of $N_{\rm AD}/N_{\infty}$, where $N_{\rm phot}$ is the total number of photons coming out from the outflowing corona and $N_{\infty}= N_{\rm phot}-N_{\rm AD}$}.


\begin{figure}
\centering
\resizebox{\hsize}{!}{\includegraphics{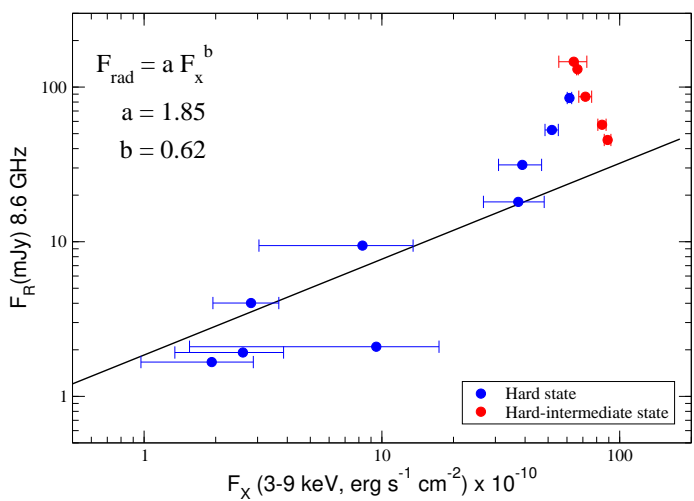}}
\caption{Radio--X-ray flux correlation for GX 339--4. The models used to compute the radio flux and the theoretical $\Gamma$ are the same as those that reproduce the time lag--$\Gamma$ correlation displayed in Fig.~\ref{lagamma}.}
\label{rad-x}
\end{figure}

\subsection{The radio -- X-ray correlation}

One of the most tight correlations in BHBs is the radio--X-ray correlation \citep{hannikainen98,corbel00,corbel03,gallo03,bright20,shaw21}. The correlation extends over five orders of magnitude in radio flux and eight orders of magnitude in X-rays. The radio -- X-ray correlation is in the form of a power-law $F_R \propto F_X^{\delta}$, where $\delta\approx0.5-0.7$ \citep{gallo12,corbel13}. In addition to the main correlation, a group of outliers populate the $F_R - F_X$ plane \citep{gallo12}. The group of outliers is associated with radio quiet sources, while most of the sources that follow the standard correlation are radio loud systems \citep{espinasse18}. 
This difference in radio emission could be an inclination effect \citep{motta18}, though other possibilities should be examined. Here we show that our model can reproduce the standard $F_R - F_X$ correlation. In particular, for the best-studied BHB GX 339--4, for which the correlation stands for more than five orders of magnitude in X-ray flux, the exponent is $\delta=0.62\pm0.01$ \citep{corbel13}. 

We follow the same procedure as in \citet{kylafis23}, where we reproduced the $F_R - F_x$ correlation for the case of a mildly relativistic outflow, i.e., $v_0 = 0.8 c$.  From the X-ray observations of the 2007-2008 outburst of GX 339-4, we obtained a relationship between X-ray flux and photon-number spectral index $\Gamma$. We rebinned the data in bins of 0.05 in $\Gamma$ for the HS and 0.1 for the HIMS. The error in the flux was computed as the standard deviation of the data in each bin. From our model, we computed the radio flux and the X-ray spectrum, i.e., the spectral index $\Gamma$, for each set of input parameters $(R_0, \tau_{\parallel})$.  Thus, we obtained an entirely theoretical relationship between radio flux and photon index $\Gamma$. We stress that our model input parameters  are not arbitrary, but they are exactly the same as those in Fig. 6 (right panel), which reproduce the correlation between time lag and $\Gamma$, shown in Fig.~\ref{lagamma} (right panel).  Thus, we attempt to explain two correlations with the same model parameters.

From the observed 
$F_X - \Gamma$ correlation and the computed $F_R - \Gamma$ one,
we matched the radio and X-ray fluxes that have the same or very similar value of $\Gamma$, and plotted one against the other.  The result is shown in Fig.~\ref{rad-x}.
We refer the reader to  Appendix A  \citep[see also][]{giannios05,kylafis23}  for the computation of the radio flux with our model as well as the details of the analysis of the X-ray observations. The black line Fig.~\ref{rad-x} is not a fit to the blue dots, but it is the observational correlation of \citep{corbel13}. The blue dots correspond to the HS and the red dots to the HIMS.  The only quantity that we had to change in this calculation, as compared with the one reported in \citet{kylafis23}, is the magnetic field $B_0$ at the base of the outflow.  Here, its value is $2.9 \times 10^4$ G.

\begin{figure}
\centering
\resizebox{\hsize}{!}{\includegraphics{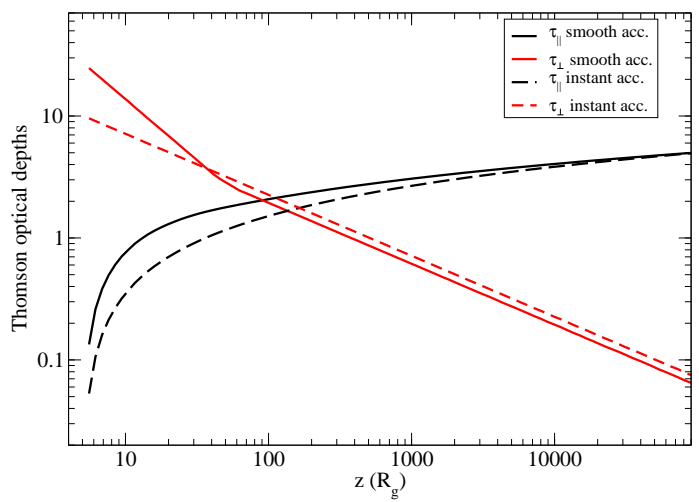}}
\caption{Thomson optical depths (parallel and perpendicular to the outflow) as  functions of distance from the black hole. We show two cases: instant acceleration of the outflow (dashed lines) and when an acceleration region is present, i.e., smooth acceleration (solid lines). We used a model with $\tau_{\parallel}=5$ and $R_0 = 100$ $ R_g$, where $R_g$ is the gravitational radius.}
\label{slab}
\end{figure}

\subsection{X-ray polarization}

Recent results from the Imager X-ray Polarimetry Explorer (IXPE) mission \citep{weisskopf22} have revealed that BHBs show polarization degrees in the X-ray band (2-8 keV) of a few percent and the polarization angle is aligned  with the outflow \citep{krawczynski22,veledina23,ingram23,rodriguez23}. Since Comptonization in a slab (with Thomson optical depth in the plane of the slab much larger than in the perpendicular direction) 
gives rise to linear polarization perpendicular to the slab  \citep{poutanen96,schnittman10}, people typically assume the Comptonizing corona to be in the form of a slab, perpendicular to the outflow, but without giving a physical justification for this. 
Also, models with a static Comptonizing region predict a lower polarization degree than the observed one. One way to produce higher polarization is by assuming that the inner disk is viewed at a higher inclination angle than the outer disk \citep{krawczynski22}. Another way is by considering an outflowing Comptonizing medium \citep{poutanen23, ratheesh23}.

Our model of a parabolic, outflowing corona naturally provides a Comptonizing, outflowing ``slab'' at its bottom.  This is shown below,  where, for simplicity of the expressions, we consider an instantaneous acceleration of the outflow to speed $v_0$ at the bottom of it.  For the full model with an acceleration region, the optical depths are calculated in Appendix A.

For a parabolic outflow, the radius of the outflow at height $z$ is
$$
R(z)= R_0 (z/z_0)^{1/2},
\eqno(1)
$$
where $R_0$ is the radius at the base of the outflow, which is at height $z_0$.  From the continuity equation
$$
2 \pi R(z)^2 n_e(z) m_p v_0 = \dot M,
\eqno(2)
$$
where $n_e$ is the electron number density, $m_p$ is the proton mass, and $\dot M$ is the mass-outflow rate, one gets for the electron density
$$
n_e(z) = n_0 (z_0/z),
\eqno(3)
$$
where $n_0$ is the density at $z_0$.
The Thomson optical depth $\tau_{\parallel}(z)$ from $z_0$ to $z>z_0$ is
$$
\tau_{\parallel}(z) = \int_{z_0}^z n_e(z) \sigma_T dz = 
n_0 \sigma_T z_0 \ln (z/z_0),
\eqno(4)
$$
while the {\it perpendicular} Thomson optical depth at height $z$ is 
$$
\tau_{\perp}(z) = n_e(z) \sigma_T R(z) = 
n_0 (z_0/z) \sigma_T R_0 (z/z_0)^{1/2}
$$
$$
\phantom{\tau_{\perp}(z) = n_e(z) \sigma_T}
= n_0 \sigma_T R_0 (z_0/z)^{1/2},
\eqno(5)
$$
or
$$
\tau_{\perp}(z) ={R_0 \over z_0}
{ {(z_0/z)^{1/2}} \over {\ln(z/z_0)} } 
\tau_{\parallel}(z).
\eqno(6)
$$

The ratio of the optical depth $\tau_{\parallel}(z)$ to the {\it total} optical depth along the outflow 
$\tau_{\parallel}(H) \equiv \tau_{\parallel}$ is
$$
{ {\tau_{\parallel}(z)} \over {\tau_{\parallel}}} =
{ {\ln(z/z_0)} \over {\ln(H/z_0)} },
\eqno(7)
$$
which means that
$$
\tau_{\parallel}(z) = { {\ln(z/z_0)} \over {\ln(H/z_0)} }
\tau_{\parallel}.
\eqno(8)
$$
Here $H$ is the height of the outflow that we take it to be $H = 10^5 R_g$, where $R_g$ is the gravitational radius.

Figure~\ref{slab} shows the variation of the parallel 
$\tau_{\parallel}$ and the perpendicular $\tau_{\perp}$ optical depths as functions of distance $z$ from the black hole, when an acceleration region is present (i.e., the physical case, solid lines, Appendix A) and when it is absent (i.e. the non-physical, but simpler, case of instant acceleration, dashed lines). 
Figure~\ref{slab} implies that, at its bottom, the outflow behaves like a slab, because $\tau_{\perp} >> \tau_{\parallel}$. In Sect.~\ref{hdis} (see also Fig.~\ref{height}) we indicated that most of the scatterings occur near the bottom of the accretion flow, especially in the HS, because the optical depth is high. 
Thus, in the HS,  the bottom of the outflow is seen by the incoming soft photons as an ``outflowing slab" and therefore the X-ray polarization is expected to be along the outflow. In the HIMS, on the other hand, the outflowing corona (at least for GX 339-4) is nearly transparent ($\tau_{\parallel} \sim 1$).  The soft photons travel along the outflow and scatter on average once.   Thus, in the HIMS, the X-ray polarization is expected to be perpendicular to the outflow.


\section{Summary and conclusion}
\label{summary}

We have demonstrated that we can reproduce most of the results from our previous
works even with a non-relativistic outflow. These are: {\em i)}  the energy
spectrum (Fig.~\ref{jetspec}), {\em ii)}  the dependence of the time-lag on
Fourier frequency (Fig.~\ref{jetlag}),
{\em iii)} the log-linear dependence of the time lag on photon energy,
{\em iv)} 
the correlation between the time lag and the photon index $\Gamma$
in GX 339--4 and Cyg X--1 (Fig.~\ref{lagamma}), {\em v)} the time-lag -- cutoff-energy
correlation observed in GX 339--4 (Fig.~\ref{cutoff}), and {\em vi)} the fact that the outflow
provides a natural lamp post for the hard X-ray photons that return to the
disk (Fig.~\ref{height}).   

The reduction of the outflow speed implies that the
fraction of back-scattered photons increases and their spectrum displays about the same $\Gamma$ as the photons directly escaping to the observer (unlike the models with $v_0= 0.8 c$, which produce softer spectra for the photons that return to the disk compared to those that go directly to the observer).
This is because the boost in the forward direction is highly reduced.
Hence the number of photons that travel at larger distances decreases.
Owing to the smaller boost along the outflow axis in the case of $v_0=0.1c$, as compared to the case of $v_0=0.8c$, there is little inclination dependence of the photon index
(Fig.~\ref{angle-index}), as expected. 
A more realistic calculation would require a parabolic outflow with a distribution of outflow velocities that decreases as one moves away from the axis. In other words, the outflow should be composed of a
mildly relativistic and narrow part at its core and a less and less relativistic
outflow at larger transverse distances.

As in the case of a mildly relativistic outflow ($v_0= 0.8 c$),  our model 
with $v_0= 0.1 c$ reproduces the observations
by changing only two parameters: the optical depth along the outflow axis $\tau_{\parallel}$ and
the radius $R_0$ at its base, and these two parameters are correlated
(Fig.~\ref{tau-R0}), so our model has only one parameter.  Our simulations in the present work ($v_0= 0.1 c$ and for GX 339--4)
require a slightly smaller range in optical depth ($1 \simless \tau_{\parallel} \simless 6$) and a slightly larger range in outflow radius ($50 \simless R_0/R_g \simless 1.5 \times 10^3$)
compared to the $v_0= 0.8 c$ case for which $2 \simless \tau_{\parallel} \simless 11$ and $30 \simless R_0/R_g \simless 600$.  The mass-outflow rate is of the order of $1 - 5$ times the Eddington rate for a 10 solar-mass black hole in the HS and $10 - 50$ times the Eddington rate in the HIMS. We remark, however, that the simplifying assumption of constant outflow speed makes the numbers unreliable.  Future calculations, with a realistic outflow speed as a function of radius, will address this issue properly.

Finally we note that in the HS (no matter what the outflow speed is), the bottom of the outflow, where most of the scatterings occur, is like a ``slab'', which produces X-ray polarization parallel to the outflow.  In the HIMS and for GX 339--4, for which $\tau_{\parallel}$ is of order unity, we predict that the polarization will be perpendicular to the outflow.

\begin{acknowledgements}
 We thank Alexandros Tsouros for offering us his code, which computes the radio emission from the outflow.   We also thank an anonymous referee for a thorough reading of the manuscript, which resulted in useful comments.
\end{acknowledgements}

\bibliographystyle{aa}
\bibliography{./bhb.bib} 

\begin{appendix}
\label{appendix}

\section{Description of the model}

\subsection{Introduction}

Our model simulates with Monte Carlo the process of Comptonization in an outflow of matter, ejected in BHBs from the hot inner flow in the vicinity of the black hole, perpendicular to the accretion disk.  Because the Bernoulli integral of the hot
inner flow is positive \citep{blandford99}, the matter cannot fall into the
black hole, hence part of the hot inner flow must escape as an outflow. In other
words, the hot inner flow is not just a static corona rotating around the black
hole, but a wind-like, ``outflowing corona''. The thin accretion disk in the accretion flow is the source of blackbody photons at the
base of the outflow.  These soft photons either escape uscattered or are scattered
in the outflow and have their energy increased, on average. This upscattering of the
soft blackbody photons produces the hard X-ray power law. Furthermore, the same 
upscattering causes an average time lag of the harder photons with respect to
the softer ones.

\subsection{Morphology}

Observational studies of the collimation of jets in active galactic nuclei (AGN) have shown that a large fraction of near-by AGN start with a parabolic outflow, which changes to a conical one further out \citep{asada12,kovalev20}. Invoking the morphological similarity between AGN and BHBs, we assume that the outflow in BHBs has a parabolic form or something close to it. Therefore, we model the radius of the outflow, as a function of height $z$ from the black hole, as

\begin{equation}
R(z) = R_0 ~ (z/z_0)^{\beta},
\label{rad}
\end{equation}

\noindent where $R_0$ is the radius of the outflow at its base, which is taken
to be at a distance $z_0$ from the black hole. We take the index $\beta$ to be $1/2$, though values close to it do not produce different results.

\subsection{Outflow acceleration}

At the bottom of the outflow, there must be an acceleration region ($z_0 \leq z \leq
z_1$), beyond which the
flow speed is constant and equal to $v_0$.  Thus, the speed of the outflowing
matter along the $z$ axis  is taken to be

\begin{equation}
v_{\parallel} (z) = 
\begin{cases}
v_0 ~ (z/z_1)^a, & z_0 \leq z \leq z_1 \\
v_0, & z > z_1 ,
\end{cases}
\label{vel}
\end{equation}
where we take $a=1/2$, though its exact values is not crucial.

\subsection{Optical depths}

If $n_e(z)$ is the number density of electrons along the outflow, mass
conservation requires that 

\begin{equation}
\dot{M} = 2 \pi R^2(z) m_p n_e(z) v_{\parallel} (z) ,
\label{mrate}
\end{equation}

\noindent where the factor 2 is for both sides of the outflow, 
$\dot{M}$ is the total mass-outflow rate, and $m_p$ is the proton mass.
The outflowing matter is considered to be consisting of protons and electrons
only. Eqs. (\ref{rad}, \ref{vel}, and \ref{mrate}) imply that the number density of electrons along the
outflow is given by

\begin{equation}
n(z) =  
\begin{cases} 
n_1 ~ (z_1/z)^{a+1}, & z_0 \leq z \leq z_1 \\
 n_1 ~ (z_1/z), & z > z_1 ,
\end{cases}
\label{dens}
\end{equation}

\noindent where $n_1$ is the number density of
electrons at $z_1$, while the density at the base of the outflow is $n_0= n_1
(z_1/z_0)^{a+1}$. 

The Thomson optical depth of the outflow along $z$ is given
by  

\begin{equation}
\tau_{\parallel} = \int_{z_0}^{H}n_e(z) \sigma_T dz
\end{equation}

\noindent where
$\sigma_T$ is the Thomson cross section and $H$ is the height of the outflow. 
Using Eq. (\ref{dens}) we find

\begin{equation}
\tau_{\parallel}= 
{ {n_1 \sigma_T z_1} \over a} \left[ ( z_1 / z_0 )^a -1 \right] +n_1 \sigma_T z_1 ln(H/z_1).
\label{taupar}
\end{equation}
Instead of $n_0$, we take $\tau_{\parallel}$ as a model parameter.  Thus, our two main parameters are $\tau_{\parallel}$ and $R_0$.  

The Thomson optical depth $\tau_{\parallel}(z)$ from $z_0$ to $z$, $z_0 <z<H$ is
\begin{equation}
\tau_{\parallel}(z) = \int_{z_0}^{z}n_e(z) \sigma_T dz
\end{equation}
\noindent or
\begin{equation}
\begin{split}
\tau_{\parallel}(z) =
\begin{cases} 
{{n_1 \sigma_T z_1} \over a} \left[ \left( z_1 \over z_0 \right)^a - \left(z_1 \over z\right)^a \right],  &z_0 \leq z \leq z_1 \\
{ {n_1 \sigma_T z_1} \over a} \left[ \left( z_1 \over z_0 \right)^a -1 \right] +n_1 \sigma_T z_1 ln(z/z_1),   &z > z_1 
\end{cases}
\end{split}
\end{equation}

\noindent while the  perpendicular Thomson optical depth at height $z$ is 

\begin{equation}
\tau_{\perp}(z) = n_e(z) \sigma_T R(z) = 
\begin{cases} n_1 \left( z_1 \over z \right)^{a+1} \sigma_T R_0 \left( z \over z_0 \right)^p,  & z_0 \leq z \leq z_1 \\
n_1 \left( z_1 \over z \right) \sigma_T R_0 \left( z \over z_0 \right)^p, & z>z_1 . 
\end{cases}
\end{equation}

\begin{table*}
\caption{Parameters of the model}
\label{modpar}
\begin{tabular}{cccl}
\hline
\hline
Symbol		        &Value	     &Units		&Parameter			 \\
\hline
\hline
$\beta$             &0.5$^{\,a}$   &--          &Index of the radial dependence with height of the outflow  \\
$\tau_\parallel$    &$1-6$         &--		    &Optical depth to electron scattering along the  axis of the outflow \\
$R_0$		        &$50-1500$     &$R_g^{\,b}$	&Radius of the outflow at its base \\
$\gamma_0^{\,c}$	&1.10--1.15	   &--		    &Lorentz factor of the electrons, when the flow is coasting\\
$kT_{BB}$		    &0.2	       &keV		    &Characteristic energy of the input soft photons   \\
$z_0$		        &5	           &$R_g$		&Distance of the bottom of the outflow from the black hole     \\
$H$                 &$10^5$	       &$R_g$		&Height of the outflow     \\
$z_1$		        &50	           &$R_g$		&Height of the acceleration zone \\
$a$		            &0.5	       &--		    &Exponent of the velocity profile in the acceleration zone \\
$R_{\rm disk}$	    &$10^3$	       &$R_{\rm ISCO}^{\,d}$ &Outer radius of the accretion disk   \\
$\cos \theta^{\,e}$ &--1 to 1      &--          &$\theta$ is the viewing angle \\
$m$		            &10	           &$\msun$	    &Mass of the black hole \\
$N_{\rm phot}$		&$10^7-10^8$   &--	        &Number of simulated photon beams  \\
$p^{\,f}$           &3             &--          &Index of the Lorentz factor power-law distribution \\
$\gamma_{\rm min}^{\,f}$ &1        &--          &Lower limit of the Lorentz factor power-law distribution \\
$\gamma_{\rm max}^{\,f}$ &500      &--          &Upper limit of the Lorentz factor power-law distribution \\
\hline
\end{tabular}
\tablefoot{\\$^a$: $\beta=0.5$ assumes a parabolic outflow \\
$^b$: Gravitational radius $R_g=Gm/c^2$. $R_g\approx1.48\times 10^6$ cm for a $m=10 \msun$ black hole\\
$^c$: $\gamma_0=1/\sqrt{1-(v_0^2+v_\perp^2)/c^2}$ \\
$^d$: Radius of the inner-most stable circular orbit. \\
$^e$: $\theta$ is the angle between the observer and the outflow axis. The code records the escaping direction of the photon in bins of 0.1 in $w=\cos\theta$, from $w=-1$ to $w=1$. Negative values indicate back scattering.\\
$^f$: For the computation of the radio emission.\\
}
\end{table*}

\subsection{Magnetic field}

For computational  simplicity, we assume the magnetic field in the outflow to be
along the  axis of the outflow, the $z$ axis.  The
$z$-dependence of the magnetic field is dictated by magnetic flux conservation to be

\begin{equation}
B(z)=B_0(z_0/z),
\end{equation}

\noindent where $B_0$ is the magnetic field strength at the base of the outflow. \\

\subsection{Lorentz factor}

The electrons are assumed to move on helical orbits around the magnetic field,
with velocity components $v_\parallel(z)$ and $v_\perp$. 
Their Lorentz factor is 

\begin{equation}
\gamma(z)=1/\sqrt{1-[v_\parallel(z)^2+v_\perp^2]/c^2},
\label{lgamma}
\end{equation}

\noindent where $v_{\parallel}(z)$ is given by Eq.~(\ref{vel}).
In the coasting region of the outflow ($z>z_1$), the Lorentz factor of the electrons is

\begin{equation}
\gamma_0=1/\sqrt{1-[v_0^2+v_\perp^2]/c^2}.
\end{equation}

It has been verified \citep{giannios05} that a power law distribution of electron 
velocities in the rest frame of the flow (see next subsection) gives nearly 
identical results.  This is because the dominant contribution to the
scatterings comes from the electrons that have the  lowest velocity (or the 
lowest Lorentz $\gamma$ factor), due to the steep power law of electron 
$\gamma$'s distribution required to explain  the overall spectrum.

\subsection{Radio spectrum}

For the computation of the radio spectrum produced by the outflow, the full distribution of electron speeds or Lorentz $\gamma$ must be taken into account.

In the rest frame of the flow, the electrons are generally taken to have a power-law distribution of Lorentz $\gamma$, namely
\begin{equation}
\label{ng}
N_e(\gamma_\text{co}) = N_0 ~ \gamma_\text{co}^{-p},
\end{equation}

\noindent from $\gamma_{min}$ to $\gamma_{max}$, where $p$, $\gamma_{min}$, and $\gamma_{max}$ are parameters of the model. Here, $\gamma_\text{co}$ is the Lorentz factor of the electrons in the \textit{\emph{co-moving frame}}.

In order to calculate the distribution of the electrons for an observer at rest, one would need to perform the transformation of the Lorentz factor from $\gamma_\text{co}$ to $\gamma$. If one assumes that the velocity of the outflow is constant throughout, then it can be shown that $\gamma$ is approximately proportional to $\gamma_{co}$, and thus Eq.~(\ref{ng}) also holds for an observer at rest. Since the acceleration region is small, the contribution of the acceleration region to the radio emission of the outflow is small.  Thus, we neglect the transformation from $\gamma_{co}$ to $\gamma$ and take the distribution of electrons for the observer at rest to be

\begin{equation}
\label{ng2}
N_e(\gamma) \simeq N_0 \gamma^{-p}.
\end{equation}

The normalization $N_0(z)$ can be calculated by integrating Eq.~(\ref{ng2}) from $\gamma_{min}$ to $\gamma_{max}$, and equating the expression with the co-moving electron density. This yields

\begin{equation}
N_0(z) = n(z) \sqrt{1-\frac{v_{||}^2(z)}{c^2}} (p-1) \gamma_{\text{min}}^{p-1}.
\end{equation}

Ignoring synchrotron self-Compton \citep[for a justification see][]{giannios05},
the equation for the transfer of radio photons in the outflow, 
in direction $\hat n$, along which length is measured by $s$, 
is given by

\begin{equation}
{{dI(\nu, s)} \over {ds}} = j(\nu, s) - a(\nu, s)I(\nu,s),
\end{equation}

\noindent where $j(\nu,s)$ and $a(\nu, s)$ are the emission and absorption coefficients respectively, and $I(\nu, s)$ is the intensity at frequency $\nu$ at position $s$. The formal solution of this equation is 

\begin{equation}   
I_\nu = \int_{s_1}^{s_2}ds j(\nu, s) \exp \left( -\int_s^{s_2} ds' a(\nu,s') \right). 
\end{equation}

For $\hat{n}$ perpendicular to the outflow axis, this is simplified to

\begin{equation}
I_\nu (z)= \frac{j(\nu,z)}{a(\nu, z)} [1- \exp\{-a(\nu,z)R(z)\}],
\end{equation}

\noindent where $R(z)$ is given by Eq.~(\ref{rad}),
since the emission and absorption coefficients depend only on $z$. The total power radiated per unit frequency per unit solid angle is thus given by 

\begin{equation}
\frac{d E_\nu}{dt d\nu d \Omega} = 2 \pi \int_{z_0}^H dz I_\nu(z) R(z). 
\label{power}
\end{equation}

Now, we need to specify the absorption and emission coefficients. Since the radio emission of the outflow is due to synchrotron emitting electrons, whose Lorentz factors follow a power law as in Eq.~(\ref{ng2}),
the absorption and emission coefficients can be calculated analytically \citep{rybicki79}. The expressions are

\begin{align}
    a(\nu,z) &= \frac{\sqrt{3}q^3}{8 \pi m_e} \left( \frac{3q}{2 \pi m_e^3 c^5} \right)^{\frac{p}{2}}N_0(z)B^{\frac{p+2}{2}}(z)\\ \nonumber
    &\times\Gamma \left( \frac{p}{4}+\frac{1}{6}\right)\Gamma \left( \frac{3p+22}{12}\right) (m_e c^2)^{p-1}\nu^{-\frac{p+4}{2}},
\end{align}

\noindent and

\begin{align}
    j(\nu,z) &= \frac{\sqrt{3}q^3 N_0(z) B(z)}{4 \pi m_e c^2 (p+1)} \\ \nonumber
    &\times\Gamma \left( \frac{p}{4}+\frac{19}{12}\right)\Gamma \left( \frac{p}{4}-\frac{1}{12}\right) \left(\frac{ 2 \pi m_e c^2}{3qB(z)}\right)^{\frac{1-p}{2}} \nu^{\frac{1-p}{2}},
\end{align}

\noindent where $q$ and $m_e$ are the absolute value of the charge and the mass, respectively, of the electron, $B(z)= B_0 (z_0/z)$ is the strength of the magnetic field at height $z$ assuming flux freezing, $B_0=B(z_0)$, and $\Gamma$ is the Gamma-function.

Integrating Eq.~(\ref{power}) over solid angles, gives the power per unit frequency radiated, 

\begin{equation}
P(\nu) \equiv \frac{d E_\nu}{dt d\nu} =  4 \pi^2 \int_{z_0}^H dz I_\nu(z) R(z). 
\end{equation}

Since there are two outflows with opposite directions to each other, an observer at a distance $d$, whose line of sight makes a 90-degree angle with the outflow axis, will measure a flux of

\begin{equation}
F(\nu) = \frac{2P(\nu)}{4 \pi d^2},
\end{equation}

\noindent where $d$ is the distance to the source. Since the observer's line of sight is taken to be perpendicular to the outflow axis, there is no need to account for a Doppler shift.

\subsection{Parameters of the model}

The main parameters of the model are the optical depth
$\tau_{\parallel}$ (see Eq.~\ref{taupar}) and the width of the outflow at its base $R_0$ (see Eq.~\ref{rad}).  Another parameter that we occasionally vary is the Lorentz factor $\gamma_0$ (see Eq.~\ref{lgamma}). These are the only parameters that we vary to obtain our results.  The ranges of variation of these parameters are shown in Table~\ref{modpar}.
The physical reason behind the importance of these three parameters is the following: a variation in optical depth is equivalent to a change in the
density at the base of the outflow. The denser the medium, the more scatterings are expected and more energetic photons will escape. Hence, $\tau_{\parallel}$  is the prime
parameter that drives the changes in the photon index $\Gamma$. A change in $R_0$ corresponds to a change in the size of the jet. The larger the medium, the longer distances the photons travel. Hence time lags are strongly affected by changes in $R_0$. Finally, an increase in $v_{\perp}$ (or $\gamma_0$) mimics the increase of the temperature in the case of thermal Comptonization.

The index $\beta$ (see Eq.~\ref{rad}) is also an important parameter because it defines the morphology of the outflow. However, in this case we fix it to $1/2$, which means that the shape of the outflow is parabolic.  

The rest of the parameters of the model are the blackbody temperature $kT_{BB}$ of the soft input spectrum from the accretion disk that enters at the bottom of the outflow, the height $z_0$ from the black hole of the bottom of the outflow, the total height $H$ from the black hole of the outflow, the height $z_1$ from the black hole at which the acceleration of the outflow stops, the exponent $a$ of the velocity profile in the acceleration region (see Eq.~\ref{vel}), the outer radius $R_{\rm disk}$ of the Shakura-Sunyaev accretion disk, the mass $m$ of the back hole in solar-mass units, the number $N_{\rm phot}$ of simulated photon beams by the Monte Carlo code, and the exponent $p$ of the distribution of electron Lorentz factors (see Eq.~\ref{ng}), along with the limits $\gamma_{min}$ and $\gamma_{max}$. None of these parameters is crucial and their values are shown in Table~\ref{modpar}.

To have good statistics in our Monte Carlo results, we combined all the escaping photons with directional cosine with respect to the axis of the outflow in a given range $\cos \theta_{\rm min} \leq \cos \theta < \cos \theta_{\rm max}$. 

\subsection{How the code works}

Photons from the inner part of the accretion disk, in the form of  a blackbody
distribution of characteristic temperature $T_{\rm BB}$, are injected at the
base of the outflow with an upward isotropic distribution. Each photon is given a
weight equal to unity (equivalently, it can be viewed as a beam of flux unity)
when it leaves the accretion  disk, and its time of
flight is set equal to zero.  The optical depth $\tau (\hat n) $ along the
photon's direction $\hat n$ is computed from the position where the photon started or scattered to the boundary of the outflow and a  fraction $e^{- \tau (\hat n)}$
of the photon's weight escapes and is recorded.  The rest of the weight of the
photon gets scattered in the outflow. If the effective optical depth in the outflow is
significant (i.e.,  $\simmore 1$), then a progressively smaller and smaller
weight of the photon experiences more and more scatterings.  When the remaining
weight in a photon becomes less than a small number (typically $10^{-8}$), we
start  with a new photon.

The time of flight of a random walking photon (or more accurately of its
remaining weight) gets updated at every scattering by adding the last distance
traveled divided by the speed of light.  For the escaping weight along a travel
direction we add an extra time of flight {\it outside} the comptonizing region in order to
bring in step all the photons (or better the fractions of them) that escape in a given
direction from different points of the boundary of the outflow.  The more a
fractional photon stays in the comptonizing region, the more energy it gains, on average, mainly from
the circular motion (i.e., $\vec{v_{\perp}}$) of the electrons.  Such
Comptonization can occur everywhere in the outflow.  Yet, a photon that random
walks high up in the outflow has a larger time of flight than a photon that random
walks near the bottom of the outflow. The optical depth to  electron scattering $\tau (\hat n) $,
the energy shift, and the new direction of the  photons after scattering are
computed using the corresponding  relativistic expressions.

Since the defining  parameters of a photon (position,  direction, energy,
weight, and time of flight) at each stage of its flight are computed, then
we can determine not only the spectrum of the radiation  emerging from the 
scattering medium and the time of flight of each escaping fractional
photon, but also the distribution of escaping heights from the outflow and the distribution of directions of escape.  To have good statistics in our Monte Carlo results, we combine all the escaping photons with directional cosine within a given range.

The time of flight of all escaping fractional photons is recorded in $N_{\rm bin}$ time bins of duration $\delta t$ s each. In this way, we can compute the number of photons that are emitted from the outflow in each time bin, and for any energy band. In other words, we can create $N_{\rm bin} \times \delta t$ s long light curves, for any energy band, which correspond to the resulting emission of the outflow in response to an instantaneous burst of soft photons which we assume enter the outflow simultaneously. 
Having created light curves in various energy bands, we can now compute delays between any pair of bands. 
Following \citet{vaughan97}, we compute the phase lag
and through it the time lag between the two energy bands as a function of Fourier frequency $\tau (\nu)=\phi/2\pi\nu$.  Then we compute the average time lag, $<t_{\rm lag}>$, in a given Fourier frequency range\footnote{Typically, $N_{\rm bin}=8192$, $\delta t=1/64$ s. The energy bands are taken to match the data results that we want to reproduce. The same applies to the frequency range where the average lags are computed.}.

The outflow velocity boosts the photons in the forward direction. Naturally, this effect is stronger as  the outflow speed $v_{\parallel}$ increases.  Regardless of the value of $v_{\parallel}$ (or $\gamma$ if one includes $v_{\perp}$), a fraction of photons are back scattered. The fraction of back scattered photons decreases as $v_{\parallel}$ increases. We distinguish the photons that hit the accretion disk after escape and those that do not. In this way, we can compute the reflection fraction as the number of photons that irradiate ("hit") the disk over the number of photons that escape in the direction of the observer. By dividing the accretion disk into radial zones, we can also compute the emissivity index $q$, which is the index of the power law 
$F(r) \propto r^{-q}$ of the radially dependent flux irradiating the disk by the outflow.

\end{appendix}

\end{document}